\documentstyle[pramana,epsf,floats]{ias}

\begin{document}
\mark{{Comparative performance ...}{V.K. Dhar, A.K.Tickoo, R. Koul and B.P. Dubey.}}
\title{Comparative performance of some popular ANN algorithms on benchmark and function approximation problems}

\author{V.K.Dhar $^\ast$, A.K.Tickoo, R.Koul. B.P.Dubey.$^+$} 
\address{Bhabha Atomic Research Centre, \\
Astrophysical Sciences Division.   \\
$^+$Electronic Instruments Services Division, \\
 Mumbai - 400 085,India. \\
$^\ast$E-mail :  veer@barc.gov.in  }
  
\keywords{Artificial Neural Network, Benchmark problems, Function approximation, Special functions}
\pacs{07.05.Mh, 7.05Kf ,2.60GF, 29.85}
\begin{abstract}
{We report an inter-comparison of some popular algorithms within the artificial neural network domain ( viz., 
Local search algorithms, global search algorithms, higher order algorithms and the hybrid algorithms) by applying 
them  to  the standard benchmarking problems  like the IRIS data, 
XOR/N-Bit parity and  Two Spiral. Apart from giving  a brief  description  of these algorithms, the results
obtained   for  the above  benchmark  problems  are presented in the  paper. 
The results  suggest  that  while  Levenberg-Marquardt algorithm yields  the lowest RMS error for the   N-bit Parity
and the Two Spiral problems, Higher Order Neurons algorithm gives the best results  for the IRIS data problem.
The best results for the XOR problem are  obtained  with the  Neuro Fuzzy  algorithm. The  above  algorithms were also 
applied for  solving  several  regression problems such  as cos(x)  and  a few  special  functions like the Gamma function,
the complimentary Error  function  and  the   upper tail cumulative  $\chi^2$-distribution function. 
The results of these  regression problems   indicate that, among  all the ANN algorithms used in the present 
study,  Levenberg-Marquardt algorithm  yields  the best  results. Keeping in view  the highly non-linear behaviour 
and  the wide  dynamic range of these functions, it is suggested that these functions  can be also considered as 
standard benchmark  problems for function approximation using artificial neural networks. }
\end{abstract}
\maketitle
\section{Introduction}
\label{1}
An artificial neural network (ANN) is an interconnected group of artificial neurons that use a mathematical model for information processing  to accomplish variety of tasks [1]. They  can be configured in various arrangements to perform a range of tasks including pattern recognition and classification [2].  While the theory and the implementation of ANN has been around for more than 50 years, it is only recently that  they have found wide spread practical applications. This is primarily  due to the advent of high speed, low cost computers that can support the rather computationally intensive requirement of an ANN of any real complexity. 
\par
Artificial neural networks  have been  used  successfully in various fields, like, pattern recognition, financial analysis, biology, engineering and so on,  because  of  their   merits  such as self-learning, self-adapting, good robustness and  capability of dealing with non-linear problems. They  have  also been  employed  extensively  in several branches  of  astronomy  for  automated  data   analysis  and  other  applications  like   star/galaxy classification,  time series  analysis (e.g  prediction of solar activity) , determination of  photometric  redshifts,   characterization  of peculiar   objects  such as  QSO's,  ultraluminous  IR galaxies[3][4].   With the increase of quantity and the distributing complexity of astronomical data,  its scientific exploitation requires a variety of automated tools, which are capable   of  performing  variety of tasks , such as data preprocessing, feature selection, data reduction, data mining  and data analysis[5].   In some recent applications the IUCAA  group (and their  collaborators) have used ANNs with   remarkable   success  for   problems   like  star/galaxy classification,  stellar spectra classification  etc.  Employing   principal  component  analysis (PCA)  for  reducing  the dimensionality of the data   and   a  multilayer  backpropagation  network based  ANN scheme, a fast  and robust method  has  been  developed  in [6]  for  classifying  a library of  optical stellar spectra for  O to M type stars.  It has been demonstrated in their work that the   PCA   when  combined  with ANN   reduces  the network  configuration  (and  hence computational time)   drastically   without  compromising on the classification accuracy.  An automated  classification  scheme  based  on the   above idea  has been also used   for  classifying    1273  stars   in the  CFLIB  data base [7]   with  an  added  advantage  that   by employing    a   generalized  PCA  technique,  the  authors  were  able    to restore   missing  data  for   300 stars.   A supervised  back-propagation  algorithm  was  used  to classify 2000 bright sources  from the   Calgary database  of IRAS (Infrared Astronomical  Satellite) spectra  into  17 predefined  classes  and   a success rate of   80$\%$ has  been  reported  by  the authors  [8]. Stellar  spectra classification   using   the probabilistic  Neural Network (PNN)  for  automated  classification of  about  5000   Solan Digital Sky  Survey (SDSS)  spectra  into  158 spectral  classes  has   also   been  performed  in [9]   with some encouraging  results.   The use of  ANN-based   technique  to develop  a  pipeline for  automated   segregation of  star/galaxies   to be observed  by the  Tel-Aviv University Ultra-Violet Experiment  (TAUVEX)  is also validated  by using  synthetic  spectra  in the  UV  region  as the  training  set  and  International Ultraviolet  Explorer (IUE) low  resolution  spectra  as the test  set [10][11].
\par
An  important  research activity in the field of  neural networks is  to compare the performance of  different ANN algorithms  on benchmark problems and to develop more efficient algorithms for solving real world problems  with noisy  and scarce data. 
It has been  also noticed  by  several workers,  that neural network algorithms are often benchmarked rather poorly [12].  More importantly,  it is also found in the literature   that   performance  of any algorithm is  only compared  to  the standard backpropagation algorithm [13]  even though  there  are  several powerful and  widely  used  algorithms   readily  available now.   Keeping this in mind, we  present  a detailed  study 
in  this paper  where  the   performance   of three generations of neural network algorithms i,e Ist order algorithms (Standard Backpropagation and Resilient Backpropagation), IInd order algorithms (Conjugate Gradient, Levenberg-Marquardt, Radial Basis Function, Simulated Annealing),  the Hybrid models like the Higher Order Neuron model and  Neuro-Fuzzy system, is examined   by applying  them to  standard  benchmarking problems  like IRIS data, XOR/N-Bit parity and  Two Spiral data.  In addition to benchmark problems discussed above,  we have also applied  the above mentioned  neural  network algorithms  for   solving  several  regression problems such  as cos(x)  and   a few  special  functions like the Gamma function, the complimentary Error  function  and  the upper tail cumulative  $\chi^2$-distribution function.  A  short  introduction to  ANN methodology and a brief  description of the ANN algorithms used  in the present  work   has also been  presented  in the paper so that  the manuscript can be  easily followed  by  researchers who are not experts in the field of  neural networks.


\section{ANN Methodology and brief description of algorithms used}
In a feed-forward ANN model, the network is constructed using layers where all nodes in a given layer are connected to all nodes in a subsequent layer. The network  requires at least two layers, an input layer and an output layer. In addition, the network can include any number of hidden layers with any number of hidden nodes in each layer. The signal from the input vector propagates through the  network layer by layer till the output layer is reached.  The output vector  represents the predicted output of the ANN and has a node  for each variable that is being  predicted.  The task of training  the ANN is to find the most appropriate set of weights for each connection  which minimizes  the output error. All  weighted-inputs are summed  at the neuron node and this summed value is then passed to a transfer (or scaling)  function.  The selection of the transfer  function is part of the neural network design and some examples of the transfer functions are  the sigmoid, hyperbolic  tangent, Sine, decaying exponential, gaussian, cauchy functions etc.  Apart  from  being   smooth and  differentiable, 
the transfer function is  chosen in such a manner  so  that it   can accept input in any range, and produce an  output in a strictly limited range. To train an ANN, initially all the neurons of the ANN are assigned random weights and the inputs and desired output vectors are presented to the ANN. The ANN uses the input vector to produce an output vector. The ANN generated output vector is compared with the desired output vector to calculate the error [14]. The ANN learns by adjusting its weights such that in the next iteration the net error produced by the ANN is generally smaller than that in the current iteration.
However,  there are several issues involved in designing and training a multilayer neural network. These are : (a)	Selecting   appropriate number of  hidden layers in the network; (b) Selecting the number of   neurons to be  used in each hidden layer; (c) Finding a globally optimal solution that avoids local minima; 
(d) Converging to an optimal solution in a reasonable period of time; (e)	Validating the neural network to test for overfitting. 
\par
Depending upon the architecture in which the individual neurons are connected and the choice of the error minimization procedure, there can be several possible ANN configurations. As discussed above, the ANN algorithms, which have been used in the present work, can be broadly categorized into three main categories viz., Local search algorithms, Global search algorithms and Hybrid algorithms.
While  algorithms like Standard backpropagation and Resilient backpropation come under the category of Local search algorithms,  Conjugate Gradient methods, Lavenberg-Marquardt algorithm,  Radial basis function and Simulated Annealing Technique belong to the  category of  Global search algorithm. Hybrid algorithm category  constitutes models like Higher order neurons and Neuro-fuzzy systems.     
A brief description of thee most promising  ANN algorithms is  presented below. 
\par
The standard backpropagation network [15]  is the most thoroughly investigated ANN algorithm. 
Backpropagation using gradient descent often converges very slowly. The success  of this algorithm in solving   large-scale problems critically  depends on user-specified learning rate and momentum parameters and, there are no standard guidelines  for  choosing  these  parameters.  Unfortunately,  if incorrect values are specified,  the convergence may be exceedingly slow, or it may not converge at all.  The  Resilient backpropagation(RProp)  algorithm was  proposed by Reidmiller [16], to expedite the learning of a backpropagation algorithm. Unlike the standard Backpropagation algorithm, RProp uses only partial derivative signs to adjust weight coefficients. The algorithm uses the so-called 'learning by epoch', which means that the weight adjustment takes place after all the patterns from the learning sample are presented to the network. 
\par
As already discussed, the learning schemes followed by  Backpropagation or the Resilient Backpropagation, based on the gradient descent methods, have several limitations. In these gradient-based  algorithms it is difficult to obtain a unique set of optimal parameters due to the existence of multiple local optima [17]. The presence of these local minima hampers the search for global minimum because these  algorithms frequently get trapped in local minima regions and hence incorrectly identify local minimum as the global minimum.  The traditional conjugate gradient algorithm uses the gradient to compute a search direction and then  a line search algorithm is used  to find the optimal step size along a line in the search direction. The Scaled conjugate gradient algorithm developed by Moller [18] is an improvement over conjugate gradient  which besides giving higher accuracy  also reduces  the number of iterations and computation time. The Levenberg  algorithm  [19] involves the use of "blending" between the steepest descent method employed by the backpropagation algorithm and  the  quadratic rule  employed in conjugate algorithms. The original Levenberg  algorithm was improved further by Marquardt, resulting in the Lavenberg-Marquardt algorthm,  by incorporating the local curvature information. In essence, the model suggests that we should move further in the direction in which the gradient is smaller in order to get around the classic "error valley".  Radial Basis Functions  are powerful techniques for interpolation in multidimensional space  and in artificial neural networks they  are utilized as activation functions [20].
Simulated annealing is a generic probabilistic algorithm for the global optimization problem, namely locating a good approximation to the global optimum of a given function in a large search space [21]. 
Starting from some random point, the error at this point $(E_A)$ is evaluated from the model or data. Then a nearby point is chosen at random and the error at this point $(E_B)$ is again evaluated. If this new point has a lower error, the process is repeated to find a still lower error point. However if it has a higher error, there is still  a chance for finding a lower error valley within the error surface. The probability of this is given by p=exp($E_A$-$E_B$)/kT  [22]. In other words "`uphill"' moves are permitted, albeit with decreasing probability  for large differences. This has the effect of managing to 'escape' from the local minima. 
\par
The limitations  of  standard backpropagation can be  overcome either by global techniques or by higher order models. Global search methods like the conjugate algorithms may reduce the architectural complexity but not the learning complexity. Higher order neuron model is the one which includes the quadratic and higher order basis functions in addition to the linear basis functions to reduce the learning complexity.  A higher order neuron model [23] has many aggregation and activation functions. The aggregation functions can be linear weighted sum (linear basis function), quadratic or higher order basis functions. Here the cross product of the input terms is added into the model where each component in the input pattern multiplies the entire input vector.  Neuro-fuzzy systems [24,25] refer to hybrids of artificial neural networks and fuzzy logic which  result  in a hybrid intelligent system that synergizes these two techniques by combining the human-like reasoning style of fuzzy systems with the learning and connectionist structure of neural networks.  

\section{Benchmarking of ANN algorithms}
The  comparative  performance  of the  ANN algorithms described above have been  studied  by applying  them to  standard  benchmarking problems  like IRIS data, XOR/N-Bit parity, Two Spiral data and Cosine(x).  While  we used  standard ANN package contained in the MATLAB software for implementing  the Lavenberg-Marquardt algorithm,  the  implementation of  other algorithms like 
Backpropogation, Resilient Backpropagation, Conjugate Gradient Method, Radial Basis Functions, Simulated Annealing, Neuro-Fuzzy etc.,  were  done   by  using  the   dedicated ANN simulator package BIKAS (BARC - IIT Kanpur  ANN  Simulator).
Written  in Java environment, this dedicated ANN package contains a variety  of neural network algorithms like the standard backpropagation, resilient , scale and self conjugate, higher order network functions, simulated annealing and radial basis methods, adaptive resonance theory algorithms, self growing networks and  fuzzy algorithms.  An exhaustive library of about  15  error minimization functions (like the conventional RMS error function, Hyperbolic square error, Minkowski error, Hubers error function, Cauchy error function  etc.)   and about  25 activation functions (like the sigmoid function, hyperbolic tan, sine, cosine, decaying exponential, Gaussian, bipolar logarithmic etc.)   are  also provided in  this package.  
\par
The training  and testing  of  all  the ANN  algorithms  used   in the present  work  has been done    on    a  Pentium P-III 700 MHz machine.   
Rigorous checks  were  also performed  at various  stages  to  ensure  that  the ANN configuration  used for a particular  problem  was  properly optimized  with respect  number of nodes in the hidden layer.  This was  done  by  monitoring the RMS error while training the  ANN. 
The RMS error employed here is defined as :
\begin{equation}                   
RMS = \frac{1}{PI}\sqrt{\sum \limits_{p=1}^{P} \sum \limits_{i=1}^{I} \left({D_{p i} - O_{p i}}\right)^2}    
\end{equation}  

where D$_{pi}$ and O$_{pi}$  are the desired and the observed values, P is  number  of training patterns and I is the number of output nodes.

The  optimized configuration  yielded a RMS error which  reduced  only  marginally when the number of nodes in the hidden layer was  increased  further, but at the cost of a much larger computation time. 
It is worth mentioning  here  that  while  the number  of  nodes  in the hidden  layer  used  varied  from  2 for XOR  to  15 for IRIS data, the number of nodes in the hidden layer was kept same  for a particular problem  in  different  ANN algorithms  to avoid any biasing towards a particular algorithm. The  training of the above networks was 
'early-stopped' to avoid any overfitting effects and this was done as soon as the RMS error reached a plateau.  A maximum of $\sim$ 2000 iterations were found to be optimum for all the  problems  which are  considered  in this  work. However, it is important to note that the number of iterations needed for the ANN to learn the input/output mapping depends on the complexity of the problem. In real world problems, e.g star/galaxy classification [6], spectra classification, primary energy estimation of cherenkov telescopes [26] etc. $\sim$ 10,000 iterations have been used.   
\subsection{IRIS problem}
Fisher [27]  introduced a benchmark dataset that contains the sepal and petal measurements of different types of iris flowers. There are 150 training samples available, each of which consists of four inputs and one output. The inputs are the measured lengths and widths of the petals and the sepals, and the output is the type of iris flower, such as Setosa, Versicolor, and Virginica. The distributions of the samples with respect to the dimensions of the sepals and petals are shown in Fig.1 for an easy visualization.
It is quite evident from Fig.1  that the classes of Versicolor and Virginica overlap, whereas the class of Setosa is clearly separated from the other two classes. 
\begin{figure}[htbp]
\epsfxsize=12cm
\centerline{\epsfbox{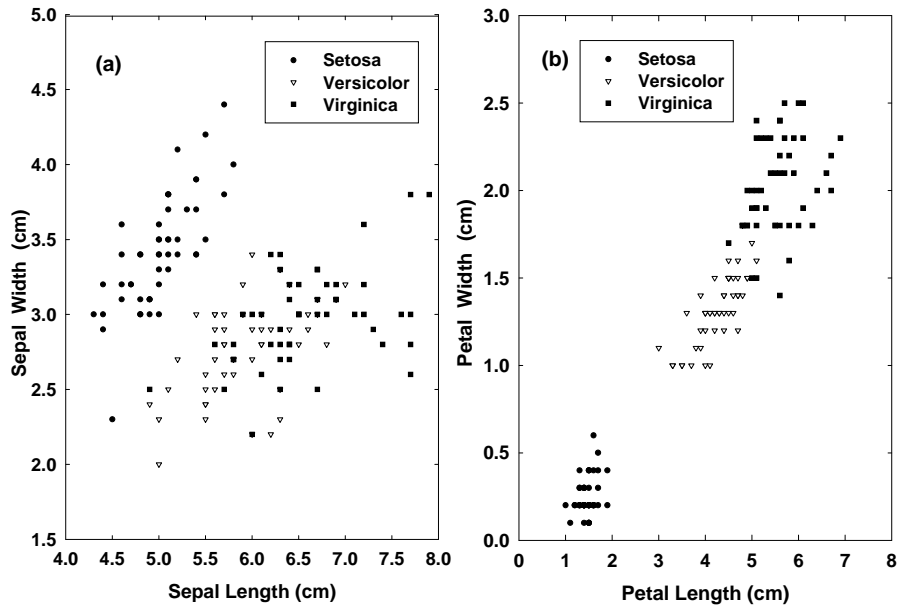}}
\caption{\label{Figure 1} IRIS data distribution with respect to the sepal and petal lengths.}
\end{figure}
In order  to  convert the training  data  in the [0,1] interval, all measurement values were first  divided by 10.  The three iris species setosa, versicolor, and virginica were categorized with the numbers 1, 2 and 3, respectively. 
The configuration  employed for training the ANN consists of 4:15:1 i.e, 4 neurons in the input layer, 15 neurons in the hidden layer and 1 neuron in the output layer corresponding to the category of the species.  The choice of 15 neurons was found to be optimum for this task. Since the final RMS error also depends upon the choice of initial parameters (like $\alpha$, $\beta$ and  the initial weights), these parameters were changed randomly  5 times  and final RMS error presented here is the mean of  these five  RMS error values.  The  RMS   error obtained  at the end of the training process  for  all the  algorithms  alongwith  time  taken  for  completing the training  is presented in Table.1. 
\begin{table}[h]
\begin{center}
\begin{tabular}{|c|c|c|c|c|c|c|c|c|}
\hline
$ANN$ & $BAC$& $RES.$ &  $LAV$ & $CON$ &   $RAD $   &   $SIM$ & $NEU$ & $HIG$\\ 
$Algorithm$ & $PRO$& $BAC$ &  $MAR$ & $GRA$ &   $BAS $   &   $ANN$ & $FUZ$ & $ORD$\\ 
\hline
$RMS $ & $2.00\times$ & $1.75\times$ & $1.92\times$ & $9.64\times$ & $3.99\times$ &  $1.20\times$ & $1.14\times$ & $1.21\times$ \\ 
$Error$& $10^{-3}$ &  $10^{-3}$ & $10^{-5}$  & $10^{-3}$ &  $10^{-3}$ &  $10^{-3}$ & $10^{-3}$ & $10^{-5}$ \\ 
\hline
$Time$  & $$ &  $$ & $$  & $$ &  $$ &  $$ & $$ & $$ \\     
$(s) $  &  $33$ &  $28$ & $70$  & $51$ &  $118$ &  $123$ & $311$ & $128$\\
\end{tabular}
\end{center}

\caption{ Mean RMS  error  and training  time of  various ANN algorithms with a configuration of 4:15:1  for the IRIS problem. The abbreviations used for different ANN algorithms are the following: BAC PRO -- Backpropagation;  RES BAC -- Resilient Backpropagation;  LAV MAR -- Lavenberg-Marquardt;  CON GRA -- Conjugate Gradient;
RAD BAS -- Radial Basis; SIM ANN -- Simulated Annealing;  NEU FUZ -- Neuro Fuzzy;
HIG ORD -- Higher Order} 
\end{table}
The test  set for the IRIS data is similar to the training set except that this data has not been presented during the training of the nets. It consists of 45 data points (15 from each class).  Instead  of  testing  the performance  of all the algorithms with test  data,   we  have only  chosen  one (or sometimes two)  ANN algorithm  for testing purpose and these are the ones  which yield  the lowest RMS  error  during  their  training  stage. Since  for the IRIS data case both  Higher Order and Lavenberg Marquardt algorithm yield  the lowest (and reasonably comparable also) RMS error, we  have used  only these  algorithms  for  checking  their  performance  on the   test data.  The test  results  obtained  for these algorithms  suggest that while  100$\%$ classification is achievable for class Setosa,  the classification for Versicolor and Virginica is only 80$\%$. The reason for not been able to obtain 100$\%$ classification between versicolor and virginica seems to be the overlapping between these two species (Fig.1). 
\subsection{XOR and N-Bit parity problems}
XOR is a standard and thoroughly investigated problem in the field of neural network research. Its popularity originates from the fact that, being able to solve it was a breakthrough achieved by back-propagation algorithm, compared to the situation faced when no learning algorithm was known to solve a non linearly separable classification task such as XOR [28]. Apart from the  XOR  problem we  also applied  the other  ANN algorithms to the generalized XOR problem i.e the N-bit parity, where the task requires to classify the sequence consisting of 1's and 0's according to whether number of 1's is odd or even [28].  The target for the net here is 1 or 0 depending on whether the sequence is odd or even. 
In the XOR problem the algorithm used has the form 2:2:1 i,e 2 neurons in the input layer,  2 neurons in the hidden  layer and 1 neuron in the output  layer.  Also, for training the networks,  more data points were also generated by incorporating  random noise of 10$\%$ at  the  XOR inputs. The  RMS  error obtained  for the XOR problem at the end of the training process,  for  all the  algorithms, alongwith  time  taken  for  completing the training  is presented in Table 2. 
\begin{table}[h]
\begin{center}
\begin{tabular}{|c|c|c|c|c|c|c|c|c|}
\hline
$ANN$ & $BAC$& $RES $ &  $LAV$ & $CON$ &   $RAD $   &   $SIM$ & $NEU$ & $HIG$\\ 
$Algorithm$ & $PRO$& $BAC$ &  $MAR$ & $GRA$ &   $BAS $   &   $ANN$ & $FUZ$ & $ORD$\\ 
\hline
$RMS$ & $1.23\times$ &  $7.72\times$ & $1.59\times$  & $6.66\times$ &  $2.70\times$ &  $1.18\times$ & $2.88\times$ & $3.67\times$ \\         
$Error$ & $10^{-3}$ &  $10^{-3}$ & $10^{-8}$  & $10^{-5}$ &  $10^{-3}$ &  $10^{-4}$ & $10^{-9}$ & $10^{-7}$ \\       
\hline
$Time$  & $$ &  $$ & $$  & $$ &  $$ &  $$ & $$ & $$ \\     
$(s)$  & $16$ &  $14$ & $15$  & $16$ &  $18$ &  $12$ & $15$ & $28$\\
\end{tabular}
\end{center}
\caption{ Mean RMS  error and  training time of  various ANN algorithms with a configuration of 2:2:1  for  the XOR problem. Full form of the abbreviations used
for different ANN algorithms can be seen in the Caption of Table 1. } 
\end{table}
As seen from Table 2, the lowest RMS  error for  the  XOR  problem is  yielded by  the Neuro-Fuzzy and the Marquardt-Lavenberg algorithms  and hence performance   testing  on  test  data sample  is done only for these two algorithms. Both these networks show $\sim$100$\%$ success rate in reproducing the XOR truth table.   
\par
The parity problem too is a demanding classification task for neural networks to solve, because the target-output changes whenever a single bit in the input vector changes. The N- bit parity consists of 2$^{N}$ (here N = 4) training pairs. A 4:2:1 architecture was used by  us  for  studying    this problem. The  RMS   error obtained  for the N-Bit  problem,  at the end of the training process,  for  all the  algorithms, alongwith  time  taken  for  completing the training  is presented in Table.3. 
The test set for N-Bit parity problem consists of 10 randomly generated noisy events (noise 10$\%$).
Testing of the net was done only with the Marquardt-Lavenberg algorithm network since compared to other algorithms  considered in this works, it gives the lowest  RMS error. The results obtained on the test data  suggest that the 4-Bit parity is reproduced with an accuracy of $\sim 90\%$. 
\begin{table}[h]
\begin{center}
\begin{tabular}{|c|c|c|c|c|c|c|c|c|}
\hline
$ANN$ & $BAC$& $RES.$ &  $LAV$ & $CON$ &   $RAD $   &   $SIM$ & $NEU$ & $HIG$\\ 
$Algorithm$ & $PRO$& $BAC$ &  $MAR$ & $GRA$ &   $BAS $   &   $ANN$ & $FUZ$ & $ORD$\\ 
\hline
$RMS$ & $9.81\times$ &  $7.12 \times$ & $3.43\times$  & $4.03\times$ &  $1.27\times$ &  $9.01\times$ & $3.42\times$ & $5.16\times$ \\         
$Error$ & $10^{-7}$ &  $10^{-7}$ & $10^{-8}$  & $10^{-7}$ &  $10^{-4}$ &  $10^{-7}$ & $10^{-3}$ & $10^{-4}$ \\     
\hline
$Time$  & $$ &  $$ & $$  & $$ &  $$ &  $$ & $$ & $$ \\  
$(s) $  & $26$ &  $40$ & $50$  & $16$ &  $28$ &  $28$ & $44$ & $55$\\
\end{tabular}
\end{center}
\caption{ Mean RMS  error and  training time of  various ANN algorithms with a configuration of 4:2:1  for  the  4-Bit parity  problem. Full form of the abbreviations used for different ANN algorithms can be seen in the Caption of Table 1.}
\end{table}
\subsection{Two-spiral problem}
The  original  two  intertwined  spirals benchmark problem was designed  by Lang and Witbrock [29]  to test  the performance of classification on binary  data. This particular task is difficult for most current algorithms since it requires the ANN model to learn the highly non-linear separation of the input space. In this benchmarking problem, two spirals, each of which has three complete turns, are created inside  a unit square Fig.2. The two-intertwined spirals problem has also been used quite extensively by other researchers  as standard benchmark  problem and  requires the neural network to learn a mapping that distinguishes between points on two intertwined spirals.
\begin{figure}[htbp]
\epsfxsize=12cm
\centerline{\epsfbox{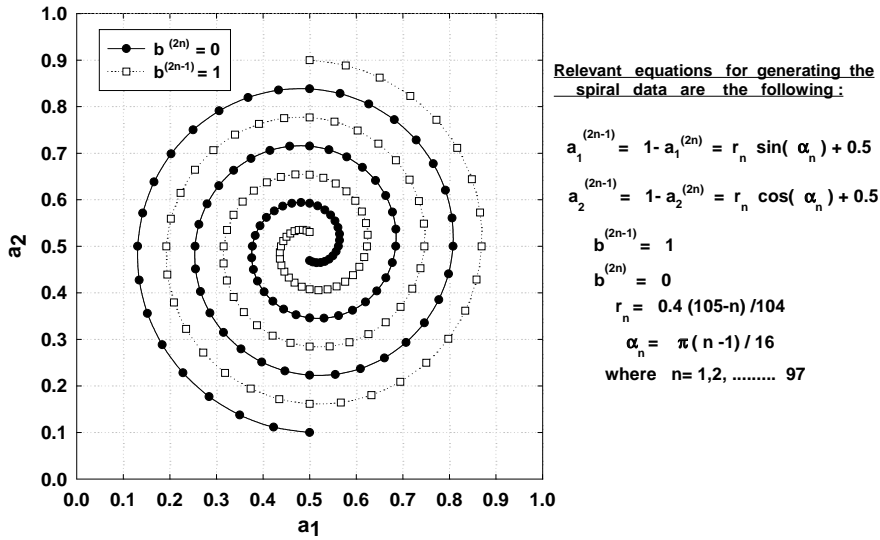}}
\caption{\label{Fig2} Distribution of data points on the two interwined spirals.}
\end{figure}
The  data used  by us for the 2-Spiral problem  contains  194  data points (97 samples per spiral).  The network configuration chosen to represent this problem has the 2:15:1 architecture,  where the two inputs correspond to a$_1$, a$_2$  values of the two spirals and the 1 output corresponds to whether the value belongs to the spiral 1 or spiral 2 ( 1 if the point falls on one spiral and 0 if it falls on other spiral). Out of these 194 data points 164 were used for training and 30 points (15 from each spiral) were used for testing. The training  results  obtained  for  all the ANN algorithms  used in the present  study  are presented in Table 4 and it is  
\begin{table}[h]
\begin{center}
\begin{tabular}{|c|c|c|c|c|c|c|c|c|}
\hline
$ANN$ & $BAC$& $RES.$ &  $LAV$ & $CON$ &   $RAD $   &   $SIM$ & $NEU$ & $HIG$\\ 
$Algorithm$ & $PRO$& $BAC$ &  $MAR$ & $GRA$ &   $BAS $   &   $ANN$ & $FUZ$ & $ORD$\\    
\hline
$RMS$ & $1.51\times$ &  $7.13\times$ & $1.09\times$  & $1.23\times$ &  $1.37\times$ &  $1.70\times$ & $1.38\times$ & $1.13\times$ \\   
$Error$ & $ 10^{-1}$ &  $10^{-1}$ & $10^{-1}$  & $10^{-1}$ &  $10^{-1}$ &  $10^{-1}$ & $10^{-1}$ & $10^{-1}$ \\  \hline
$Time$  & $$ &  $$ & $$  & $$ &  $$ &  $$ & $$ & $$ \\   
$(s)$  & $60$ &  $56$ & $110$  & $160$ &  $315$ &  $200$ & $390$ & $190$\\    
\end{tabular}
\end{center}
\caption{ Mean RMS  error and  training time of  various ANN algorithms with a configuration of 2:15:1 for  the 2-Spiral  problem. Full form of the abbreviations used
for different ANN algorithms can be seen in the Caption of Table 1.}
\end{table}
quite evident  from  this  table  that  the  Levenberg-Marquardt gives the best convergence  results.  However, from  training  time considerations,
the standard backpropagation and the resilient backpropagation  algorithms are seen to consume  minimum  training  time. Performance  check of the  Levenberg-Marquardt algorithm,   on  test data for 30 points (15 from each spiral),   indicates   that  $\sim$ 70 $\%$ of the events are classified in the proper spiral category. This is much better as compared to   resilient  backpropagation  which  can classify only $\sim$ 50$\%$ of the events properly.
\par
A consolidate  report  of  the   mean  RMS  error  yielded  by   various   ANN algorithms  used  in the present  work   for  all  the   4   benchmark  problems  is  presented  in  Table 5.  A  plot  of  the RMS error  as a function of  number  of  iterations,   for   all the   4  benchmark  problems  is  also shown  in  Fig. 3  so  that the performance  of the  backpropagation   algorithm   can be   compared   with   other   algorithms.
\begin{table}[h]
\begin{center}
\begin{tabular}{|c|c|c|c|c|}
\hline
 $ANN$       & $IRIS$    & $XOR $    &  $N-bit$  & $2-spiral$ \\ 
$Algorithms$ & $problem$ & $problem$ &  $parity$ & $problem   $ \\ 
\hline
$Backprop$ & $2.00\times10^{-3}$ &  $1.23\times10^{-3}$ & $9.81\times10^{-7}$  & $1.51\times10^{-1}$ \\         
\hline
$Resilient$ & $1.75\times10^{-3}$ &  $7.72\times10^{-3}$ & $7.12\times10^{-7}$  & $7.13\times10^{-1}$ \\       
\hline
$Lavenberg$  & $1.92\times10^{-5} $ &  $1.59\times10^{-8}$ & $3.43\times10^{-8}$  & $1.09\times10^{-1}$  \\     
\hline
$Conjugate$ & $9.64\times10^{-3}$ &  $6.66\times10^{-5}$ & $4.03\times10^{-7}$  & $1.23\times10^{-1}$ \\         
\hline
$Rad. Basis$ & $3.99\times10^{-3}$ &  $2.70\times10^{-3}$ & $1.27\times10^{-4}$  & $1.37\times10^{-1}$ \\         
\hline
$Sim. Annealing$ & $1.20\times10^{-3}$ &  $1.18\times10^{-4}$ & $9.01\times10^{-7}$  & $1.70\times10^{-1}$ \\         
\hline
$N. Fuzzy$ & $1.14\times10^{-3}$ &  $2.88\times10^{-9}$ & $3.42\times10^{-3}$  & $1.38\times10^{-1}$ \\         
\hline
$Hig. Order$ & $1.21\times10^{-5}$ &  $3.67\times10^{-7}$ & $5.16\times10^{-4}$  & $1.13\times10^{-1}$ \\         
\end{tabular}
\end{center}
\caption{ Comparison of mean RMS error for the different ANN algorithms considered for the study of benchmark problems. } 
\end{table}
For the sake  of  clarity,  the  RMS  error  is  shown  only   for  the  Backpropagation  algorithm   and   one  more   specific  algorithm   which  yields   the minimum  RMS error (i.e    Higher Order for IRIS,  Neuro-fuzzy for XOR and Levenberg-Marquardt method for N bit parity and 2 spiral problem).  
\begin{figure}[htbp]
\epsfxsize=11cm
\centerline{\epsfbox{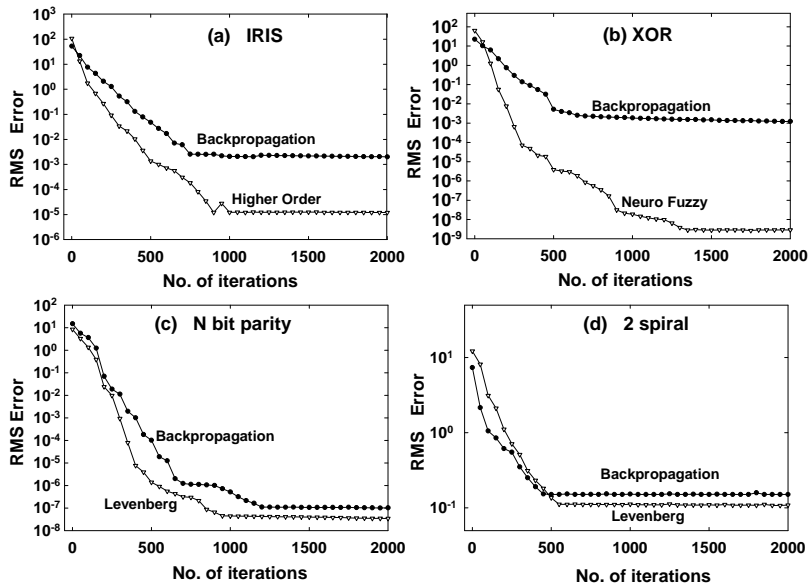}}
\caption{\label{Fig3} Variation of RMS error as a function of number of iterations  for the  4 benchmark problems: (a) IRIS (b) XOR  (c) N Bit Parity  (d) 2 Spiral. The  RMS  error  is  shown  only   for  the  Backpropagation  algorithm   and   one  more   specific  algorithm   which  yields   the minimum  RMS error  for a particular problem.}
\end{figure}
It is evident  from Fig.3 and Table 5 that the Lavenberg-Marquardt  method  yields   a  lower   RMS error   as   compared to the standard  backpropagation method.   Even  for   the    IRIS  and  XOR   problems,   where  Higher Order  and  Neuro Fuzzy algorithms, respectively   are   found   to   yield   the   lowest  RMS  error,  the  performance  of   the   Levenberg-Marquardt    is   always    better  than  the  remaining    ANN  algorithms.   The   reason for  the  superior   performance  of   Levenberg-Marquardt    is  due  to  the  fact    that  it is a combination of  gradient decent and  Gauss-Newton method   which   combines   the advantages  of the local convergence  properties  of the   Gauss-Newton   method  and the global  properties of gradient  descent.  However,    the computational   complexity of backpropagation  is  only  O(n)  as against  O(n$^3$)   for the   Levenberg-Marquardt  algorithm (where  n is the   total  number of  weights  in the network).  
\section{Application  of ANN algorithms  to regression  problems}
Artificial  neural networks   have  become a popular  tool for approximating non-linear functions  in  higher  dimensions.  Although they are not the panacea for these type of problems, they are nevertheless recognized as a useful   tool  for approximating non-linear  functions.  Other well known methods  which are conventionally  used  for  these  problems include  splines [30], additive models [31], MARS [32], hinging hyperplanes [33] and  CART [34].  While none of  these  methods are likely to perform consistently  better than the others across a wide range of problems,  it  is indeed a non-trivial  task to develop a method  that is truly effective  for  all types of non linear  functions.   Keeping  in view  the superior ability of ANNs to capture non-linear behaviour of a function and its reasonably  fast computation speed,  we were tempted to apply the ANN as a regression tool  for  approximating  functions like cos(x) and  a few special functions  like  the Gamma function, the complimentary Error function and the  upper tail cumulative  $\chi^2$-distribution function [35].
\subsection{Approximation  to cos(x)}
To  test  the performance  of  the ANN algorithms  as a regression analysis tool,  we have first  applied the  ANN  algorithms to a simple trigonometric  function like  y=cos(x). 
In order to keep  the output range of the network between 0 and 1,  we follow  the approach given in [28] where the function is changed to y =(cos(2x)+1)/3. The ANN configuration chosen  for this problem 
(i.e, 1:2:1) and  the number  of  data points  used for training (=200) is  again  similar to  that used  by [28]. The training data  set  for this  problem   is  synthesized  by evaluating  the  function  y =(cos(2x)+1)/3  at  200 randomly  chosen  points  which are picked  uniformly  in the interval [0,$\pi$] range (Fig. 4a). Additional  100 data points, following the same prescription,   were also  generated   for testing  the  best  ANN  algorithm  which produces the  lowest  RMS  error  during training.  The training  results  obtained  for  all the ANN algorithms used in the present  study  are presented in Table 6. 
\begin{table}[h]
\begin{center}
\begin{tabular}{|c|c|c|c|c|c|c|c|c|}
\hline
$ANN$ & $BAC$& $RES.$ &  $LAV$ & $CON$ &   $RAD $   &   $SIM$ & $NEU$ & $HIG$\\ 
$Algorithm$ & $PRO$& $BAC$ &  $MAR$ & $GRA$ &   $BAS $   &   $ANN$ & $FUZ$ & $ORD$\\      
\hline
$RMS$ & $9.83\times$ &  $7.61\times$ & $3.29\times$  & $7.30\times$ &  $4.15\times$ &  $4.61\times$ & $6.71\times$ & $8.12\times$ \\     
$Error$ & $10^{-6}$ &  $10^{-5}$ & $10^{-7}$  & $10^{-6}$ &  $10^{-5}$ &  $10^{-6}$ & $10^{-6}$ & $10^{-7}$ \\   \hline
$Time$  & $$ &  $$ & $$  & $$ &  $$ &  $$ & $$ & $$\\
$(s)$  & $7$ &  $7$ & $12$  & $8$ &  $12$ &  $15$ & $25$ & $20$\\ 
\end{tabular}
\end{center}
\caption{ Mean RMS  error and  training time of   various ANN algorithms with a configuration of 1:2:1 for  the  (cos(2x)+1)/3  problem. 
The abbreviations used for different ANN algorithms are the following: BAC PRO -- Backpropagation;  RES BAC -- Resilient Backpropagation;  LAV MAR -- Lavenberg-Marquardt;  CON GRA -- Conjugate Gradient;
RAD BAS -- Radial Basis; SIM ANN -- Simulated Annealing;  NEU FUZ -- Neuro Fuzzy;
HIG ORD -- Higher Order.} 
\end{table}
The results  of the generalization performance  of the Levenberg-Marquardt algorithm, which yields the lowest RMS error during  training, is shown in Fig.4.
\begin{figure}[htbp]
\epsfxsize=10cm
\centerline{\epsfbox{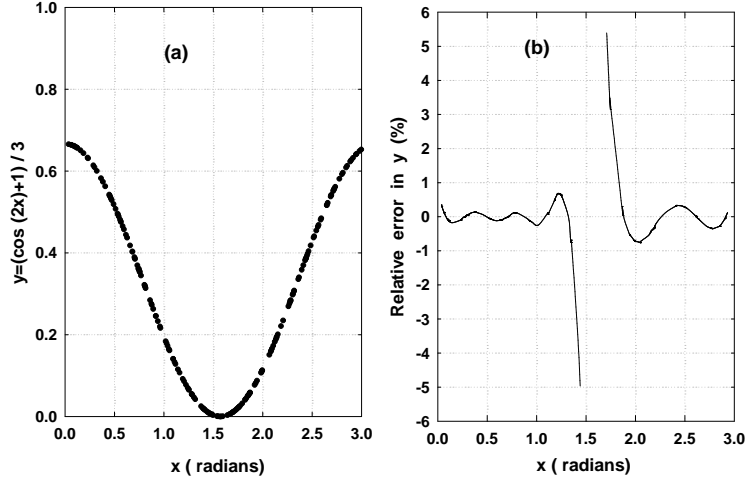}}
\caption{\label{Fig4} (a) The training data set for the function y =(cos(2x)+1)/3. (b)Performance of the  Levenberg-Marquardt algorithm in terms  of  relative  error  in approximating the function,  y =(cos(2x)+1)/3.}   
\end{figure}
In this figure,  we have plotted the  relative  error in y  (defined as  (y$_{ANN}$-y$_{EXP}$)/y$_{EXP}$)  as  a  function of  x  for 100 random data points generated  uniformly  in the interval [0,$\pi$]. 
Here, for  a given  value of x,   y$_{ANN}$ is the  value  predicted  by the  ANN and y$_{EXP}$ is  expected value of the function (cos(2x)+1)/3. 
It is evident  from Fig.4b that except for x values in the vicinity of $\pi$/2 radians, where the function  y =(cos(2x)+1)/3 itself  becomes close to zero, the relative error in y is in general  $<$1$\%$ for all other values of  x.   
\subsection{Approximation  to  a few special functions}
In this section, we  apply the ANN algorithms  as a function approximation   tool to  few special functions  like  the Gamma function, the complimentary Error function and the   upper tail cumulative  $\chi^2$-distribution function. The training and test data sets for the above special functions  have been generated by using the MATHEMATICA software package. 
\par
The  Gamma-function  ( $\Gamma(z)$) has one argument and is defined by the following integral:
\begin{equation}
\Gamma(z) = \int_0^\infty t^{z-1} e^{-t}\,\mathrm{d}t
\end{equation}
The  approximation of the  Gamma function was implemented with an ANN  configuration of 1:20:1, where  the one node in the input corresponds to  the z value  in the         
range  0$<$z$<$20 and the output 
node corresponds  to  $\ln$ $\Gamma(z)$.  The purpose  of  using    $\ln$ $\Gamma(z)$  
instead  of  $\Gamma(z)$  directly  was to avoid  overflow  problems  even  at a quite modest value of z.   The training of the ANN algorithms has been done with $\sim$1000 values and only those  values of z and  ln$\Gamma(z)$ are used for which 
$\Gamma(z)$ $<$ 1.2$\times$10$^{17}$.
\par
The  second special function chosen by us to test the function approximation capability of the ANN is the complimentary function, erfc(x). The complimentary error function has one argument and is defined  by the following integral:
\begin{equation}
erfc(x)=  \frac{2}{\sqrt{\pi}} \int^{\infty}_{x} \; e^{-t^2} \;dt 
\end{equation}
Since  there is direct relationship between between  the complimentary error  function and the cumulative distribution for the Gaussian distribution, we have tried to apply the ANN algorithms for approximating the normal tail integral. The upper tail integral, or the cumulative upper distribution function, Q(x) for Gaussian probability distribution with argument x is defined by 
\begin{equation}
Q(x)=  \frac{1}{\sqrt{2\pi}} \int^{\infty}_{x} \; e^{-t^2/2} \; dt=  \frac{1}{2} \;erfc( \frac{x}{\sqrt{2}}) 
\end{equation}
The function approximation for  the  normal tail probability integral was implemented with a ANN configuration of 1:20:1, where the one input node corresponds to the x value ranging between 0 to 20 and the  output node corresponds to ln Q(x). The  values  of  Q(x)  are  in the range  $\sim$2.767 $\times$10$^{-89}$  to 0.5.
About 1000 values  of x  and  (ln Q(x))  were used for training  the ANN algorithms. 
\par
The third special function chosen for testing the function approximation capability of the ANN  is the 
cumulative distribution function of  the  $\chi^2$-probability  distribution.  The chi-square distribution is one of the most widely used theoretical probability distribution  in inferential statistics, i.e. in statistical significance tests.  The best known situation in which the   $\chi^2$-distribution  is  used are the common  $\chi^2$-tests for goodness of fit of an observed distribution to a theoretical one.   The   $\chi^2$-upper tail cumulative distribution function ( Q($\chi^2$$|$$\nu$)) is defined by the following integral:
\begin{equation} 
Q(\chi^2|\nu)= \frac{1}{2^{\nu/2}\,\Gamma(\nu/2)}\ \int^{\infty}_{\chi^2} e^{-t/2} \; t^{\nu/2\,-\, 1} \; dt \; ; \quad \quad \textrm{for} \; \nu>0, \;\chi^2\ge 0
\end{equation}

Where $\nu$ is the degrees of freedom.
The  approximation of the  $\chi^2$  upper tail cumulative distribution function ( Q($\chi^2$$|$$\nu$))   was implemented with  a  ANN configuration of  2:20:1 where the two input nodes correspond to the $\chi^2$ and $\nu$ values. The  output node  of the  ANN represents   the   
(ln Q($\chi^2$$|$$\nu$)) value.  About 1000 values with  1$\leq$ $\chi^2$ $\leq$ 100 and   1$\leq$ $\nu$ $\leq$ 100   were used  for training the ANN algorithms.  The training  of the ANN was performed  with only those values of $\chi^2$ and  $\nu$ which  yield  a 
(Q($\chi^2$$|$$\nu$)) between  $\sim$1.757 $\times$10$^{-23}$  to 0.999. 
The  results of the training  regarding  the mean error for all  the  three  special  functions   discussed   above  are presented  in Table 7. 
\begin{table}[h]
\begin{center}
\begin{tabular}{|c|c|c|c|}
\hline
$   ANN       $ &    RMS error     &   RMS error        &   RMS error       \\  
$Algorithm    $ &    (Gamma        &   (Upper  tail     &  (Upper tail      \\  
$Studied      $ &    Function)     &   Normal dist.)    &   $\chi^2$ dist.) \\  
\hline
$Backpropagation$ &  $ 2.25\times 10^{-2}$ &  $ 9.16\times 10^{-1} $ &  $2.91\times 10^{0} $\\    
\hline
$Resilient Backprop$ &  $8.97\times 10^{-2}$ &   $1.51\times 10^{-1}  $ &  $ 2.01\times 10^{-3} $\\         
\hline
$Lavenberg-Marquardt$ & $1.25\times 10^{-6}$ &   $2.08\times 10^{-9}  $ & $1.72\times 10^{-5}$\\     
\hline
$Conjugate Gradient$ &  $5.01\times 10^{-3} $ &  $ 2.14\times 10^{-2}   $   & $ 4.87\times 10^{-2} $\\
\hline
$Radial Basis$  &  $5.68\times 10^{-3} $ &   $ 7.33\times 10^{-3}$  & $ 5.71\times 10^{-2} $ \\       
\hline
$Simulated Annealing$  &  $ 5.23\times 10^{-3}$ &   $1.52\times 10^{-2}$  &  $ 4.20\times 10^{-3} $ \\
\hline
$Neuro-Fuzzy$ &  $4.12\times 10^{-2} $ &  $8.96\times 10^{-2}$ &  $1.16\times 10^{-3}$ \\
\hline
$Higher-Order$  &  $ 7.16\times 10^{-5} $ &  $ 9.86\times 10^{-8}$ &  $6.42\times 10^{-5} $ \\ 
\end{tabular}
\end{center}
\caption{ Mean RMS  Error of  various ANN algorithms with a configuration of 1:20:1 for  function approximation of  3 special functions.} 
\end{table}
\par
Performance of the  Levenberg-Marquardt algorithm in terms  of  relative  error  in approximating the three  special  functions  is  shown in Fig.5. 
\begin{figure}[htbp]
\epsfxsize=10cm
\centerline{\epsfbox{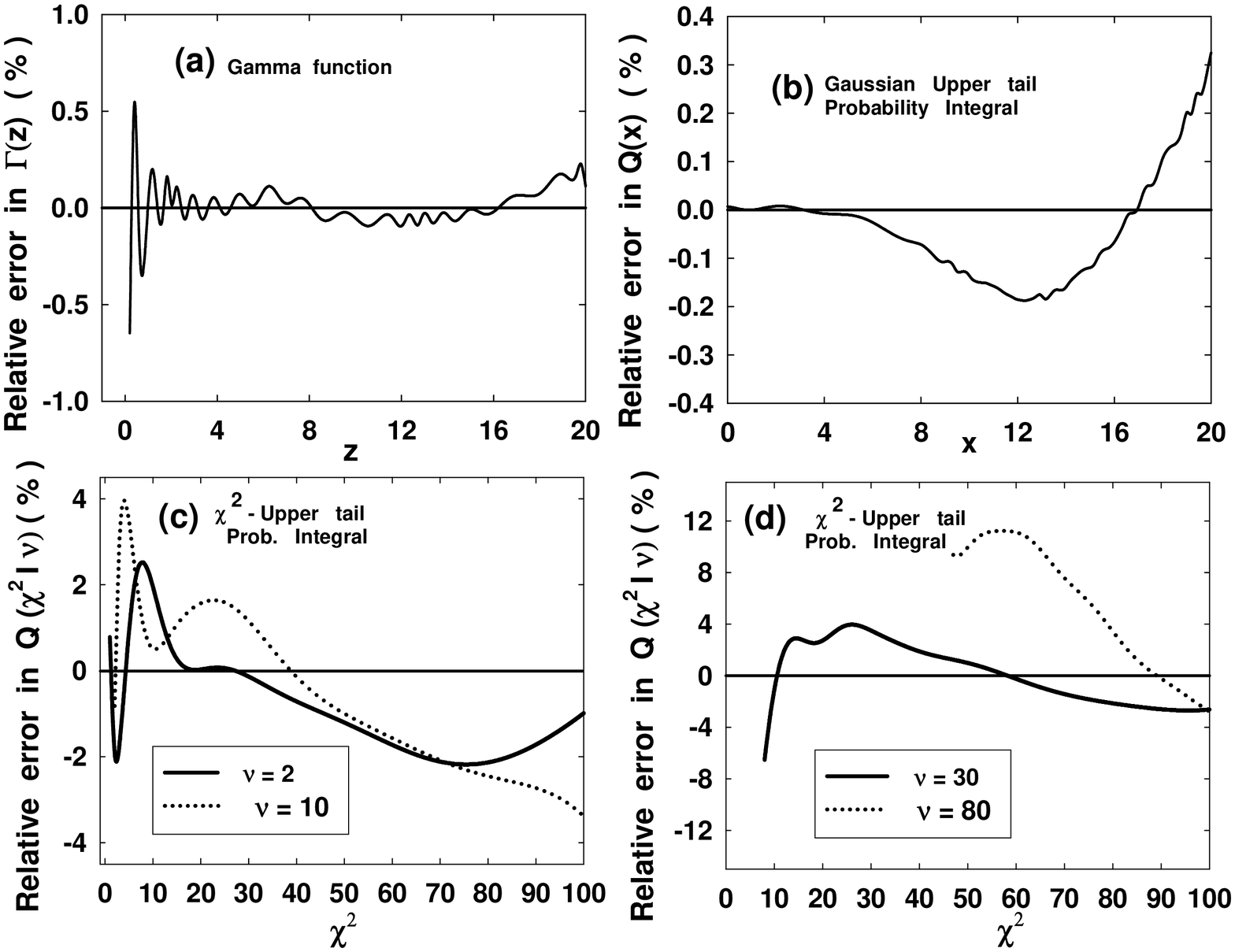}}
\caption{\label{Fig5} Performance of the  Levenberg-Marquardt algorithm in terms  of  relative  error  in approximating the following  special functions :(a) Gamma function, (b) Upper  tail probability of   Normal distribution, (c) and (d)  Upper  tail probability of   $\chi^2$ distribution for   different  values of  $\nu$.}   
\end{figure}
A data  sample of 100 values  each is used for testing the approximation for the Gamma function (Fig. 5a) and Gaussian upper tail probability integral (Fig. 5b). The corresponding data samples used for testing the   $\chi^2$ upper tail probability integral is   400  for  4 different values of   $\nu$.  Referring first to the approximation of the Gamma function, it is evident from Fig. 5a  that the relative error in $\Gamma(z)$ is $<$0.25$\%$  for  2$<$z$<$20.   However,  for  0$<$z$<$2,  the relative error  increases significantly to  $>$$\pm$0.5$\%$.  Regarding the approximation of the Gaussian upper tail probability integral, the relative error in Q(x) is within   $\pm$0.3$\%$ for all values of x in the range 0 to 20.   The results for the approximation of  $\chi^2$ upper tail probability integral (Fig.5c and Fig.5d) indicate that the relative error in Q($\chi^2$$\mid$$\nu$)  is in general significantly more than that of the other 2 special functions possibly  because of the presence of 2 input parameters ( $\chi^2$ and $\nu$)  instead of 1 as  in the case of other two special functions. Keeping in view the fact that  these special functions are being approximated over a very wide dynamic range ($\sim$ 0.88 to  
1.2$\times$10$^{17}$ for Gamma function,  $\sim$2.7 $\times$10$^{-89}$ to 0.5 for   
Gaussian upper tail integral and $\sim$1.8 $\times$10$^{-23}$ to 0.999 for   
$\chi^2$  upper tail integral), we believe that the results obtained are rather encouraging.
However, there is a strong  need to further improve these results  if one demands that  approximation using ANN algorithms should  yield a performance which is  comparable to that of the conventional methods using numerical algorithms or other ad hoc approximations. Furthermore, keeping in view the  widespread use of these functions
and also their highly non-linear behaviour  with a very wide dynamic range, we feel that  these functions can be considered as standard benchmark problems for function approximation studies  using  ANNs.       
\section{Discussion and  conclusion}
Artificial  neural network   algorithms  have been applied to a variety of problems in various  diverse areas of physics, biology, medicine, computer research etc. The main aim of most of these studies has been to use ANN-based algorithms (generally standard backpropagation) as an  alternative method  to conventional analysis  for achieving  better results.  While  comparative performance of  some  ANN algorithms  like standard backpropagation, fuzzy logic, genetic algorithms, fractals etc., has been studied  for various  applications, a rigorous  intercomparison of  some of powerful algorithms ( e.g. the ones studied in this work)  is still  missing from the literature.  The primary aim of this  work  has been to provide a rigorous comparative study of various  powerful algorithms, by first applying them to standard benchmark problems  and  then apply them  as a regression tool  for  approximating  functions like cos(x) and  a few special functions. Our  results  suggest  that  while  Levenberg-Marquardt algorithm yields  the lowest RMS  error for the   N-bit Parity and the Two Spiral problems, Higher Order Neurons algorithm gives the best results  for the IRIS data problem. The best results for the XOR problem are  obtained  with the  Neuro Fuzzy  algorithm.  It is worth mentioning  here  that  benchmark  problems ( IRIS, XOR/N-Bit Parity and 2-Spiral)  have been also studied by numerous other workers.  For example, using a 2:2:1 configuration for the XOR problem,  Wang [28] has reported that one can achieve an accuracy of  $\sim$80$\%$ with $\sim$5000 epochs of training. Other benchmark problems like Parity and cos(x)  has been also studied by the same author, but comparison is done only for the backpropagtion and simulated annealing techniques. Likewise, the 2-Spiral problem has been studied by several workers  using different algorithms like  vanilla backpropagaion [36] with configuration of 2:5:5:1, generalised regression model [37], vector quantization method [38], input coding   
scheme [39].  Complicated ANN configurations like 2:20:20:1 for the 2-Spiral problem with 50000 training epochs   and 4:4:2:1 for the IRIS problem  with 30000 training epochs has also been attempted  by   Lee [40] for achieving  reasonably accurate results  for these benchmark problems.
\par
Regarding  application of  neural  network algorithms  for  solving   regression problems,  such as  evaluation of special  functions like the Gamma function, the complimentary Error  function  and  the   upper tail cumulative  $\chi^2$-distribution function, we believe that such  an  attempt  has been made  in this work for the first time. The results obtained  in this work  indicate that, among  all the ANN algorithms used in the present  study,  Levenberg-Marquardt algorithm  yields  the  best  results.
Conventionally,  two groups of approximations are found in the literature which are used for  calculating these special functions.   One group consists of standard numerical algorithms  which, at least in theory, allow computation of the above integral to arbitrarily high  precision. However, computation using  these numerical  algorithms requires massive computation.  The  second group consists of so called "ad hoc  approximations"  which require  only a few carefully chosen numeric constants. Importantly, a serious limitation of most of the approximations in both the groups is that they are designed to work  in a predefined range of input parameter  values and the accuracy of the approximation rapidly   deteriorates   when  the input parameters  takes a  value outside  the predefined  range. 
\par
In order to appreciate the complexity of evaluating the special functions studied in this paper, it is worth discussing  here  some of important approximations used for these functions.  A well known  method  for calculating the gamma function is the so called  Lanczos approximation [41] which computes the value of $\Gamma(z)$  any positive real argument(z)  with a high level of accuracy. 
Likewise, a compilation of useful approximations used for evaluating 
the upper tail integrals for the Gaussian and $\chi^2$ distributions  can be found in [42] and [43], respectively. 
\par
Although the comparative performance  of different ANN algorithms is in general problem dependent, we feel that the  study undertaken in this paper  does  give an insight into the power of various  powerful ANN algorithms.  Since for real world problems it is not an easy task to  identify the most  suitable  ANN algorithm by just having a look at the  problem, our results  suggest  that  while  investigating the comparative performance  of  other ANN algorithm,  the  Levenberg-Marquardt algorithm  deserves  a  serious   consideration  and  cannot  be  rejected  outright  because  of  its  training  time  overheads.
\section{Acknowledgements}
We would like to thank all the members  of the BIKAS (BARC - IIT Kanpur  ANN  Simulator;
BRNS- 2000/36/5-B) collaboration  for useful discussions and suggestions.  We would  also like  to   thank    the  anonymous  referee  his  valuable  comments and suggestions  for improving the paper. 

\end{document}